\documentclass[a4paper,preprint,12pt]{elsarticle}
\usepackage{amsmath,amsthm,amssymb}
\usepackage{graphicx}
\usepackage[left=1.0in, right=0.75in, top=0.8in, bottom=0.8in]{geometry}
\usepackage{lineno}

\journal{IEEE Photonics 2011}

\begin{document}

\begin{frontmatter}

\title{Polarization sensitive solar-blind detector based on $a$-plane AlGaN.}

 \author[]{Masihhur R. Laskar}
 \author[]{A. Arora}
 \author[]{A. P. Shah}
 \author[]{A. A. Rahman}
 \author[]{M. R. Gokhale}
 \author[]{Arnab Bhattacharya}

\address[]{Department of Condensed Matter Physics and Materials Science,\\ Tata Institute of Fundamental Research, Homi Bhabha Road,
Mumbai 400005, India.}

\vspace{-0.5in}
\begin{abstract}
We report polarization-sensitive solar-blind metal-semiconductor-metal UV photodetectors based on $(11\bar{2}0)$ $a$-plane AlGaN. The epilayer shows anisotropic optical properties confirmed by polarization-resolved transmission and photocurrent measurements, in good agreement with band structure calculations.
\end{abstract}

\end{frontmatter}


Solar blind UV (SBUV) detectors, with no photosensitivity above $280$nm wavelength, have wide range of applications like -- missile plume detection, UV astronomy, chemical/biological battlefield reagent detection etc.$^{1-3}$. The wide-bandgap, high-temperature compatible AlGaN material system has been the workhorse for such SBUV detectors with many reports on high performance devices based on $[0001]$ $c$-plane AlGaN layers. The inherent anisotropic optical properties and reduced crystal plane symmetry of ``non-polar" $(11\bar{2}0)$ $a$-plane AlGaN epilayers allows the fabrication of polarization sensitive (PS) detectors. Such PS detectors give additional advantages of selectivity and narrow band detection in a differential configuration consisting of two or four photo-detectors, without using filters $^{4,5}$. We present, to the best of our knowledge, the first report of a PS SBUV detector.

\begin{figure}[!b]
\centerline{\includegraphics*[width=9cm]{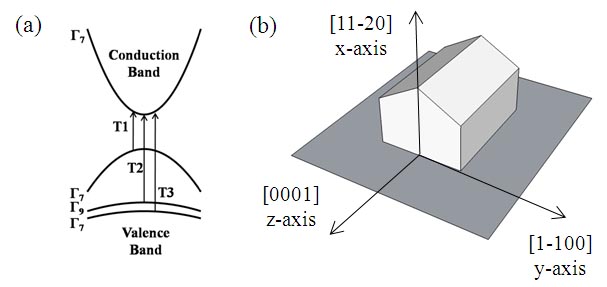}}
\caption{(a) Schematic diagram showing the three closely spaced valence band at $k$=0 of the III-nitrides. (b) Orientation of hexagonal unit cell for $a$-plane nitrides. The in-plane strains are $\epsilon_{yy}$ and $\epsilon_{zz}$.}
\end{figure}

About $0.5 \mu m$ thick Al$_{0.6}$Ga$_{0.4}$N epilayers were grown on AlN buffer layers via metal organic vapour phase epitaxy (MOVPE) in a closed-coupled showerhead reactor using standard precursors. The details of the growth procedure, method to estimate the solid phase Al content and strain in the layer can be found in Refs.[6,7]. Metal-Semiconductor-Metal (MSM) type devices with interdigitated finger geometry Schottky contacts (metallization--$200{\AA}$ Ni/$1000{\AA} $Au) were fabricated using standard optical photolithography, electron-beam evaporation and lift-off techniques.

The III-nitride semiconductors have three closely-spaced valence bands near the center of the Brillouin-zone ($k$$=$$0$) as shown in Fig.1(a). The transition probabilties of electrons from each valence band to the conduction band are different and are strongly determined by the polarization of light. For $(11\bar{2}0)$ $a$-plane epilayers, the in-plane strains are $\epsilon_{yy}$ and $\epsilon_{zz}$ as shown in Fig.1(b). Using HRXRD we estimate the in-plane anisotropic strain in our Al$_{0.6}$Ga$_{0.4}$N epilayer as $\epsilon_{yy}$=$-0.5$\% and $\epsilon_{zz}$=$+0.2$\%, for which $E1$ transition is strongly \textbf{z}-polarized and $E2$ transition is strongly \textbf{y}-polarized, obtained from the band structure calculation by solving the \emph{Bir-Pikus} Hamiltonian $^{8,9}$.

Fig.2(a) shows the absorption spectra of Al$_{0.6}$Ga$_{0.4}$N for two different polarizations, where the extrapolation of $\alpha^2$ vs. \emph{energy} plot gives the bandgaps of the epilayer as $\sim$$4.67$ eV and $\sim$$4.73$ eV for $E \parallel c$ and $E \bot c$ polarization directions respectively. So the valance band splitting $E_4$=$E2$-$E1$ is $\approx 60$ meV. Fig.2 (a) inset shows the calculated $E_4$ as a function of in-plane strain and the the black dot represents the strain in the layer. The experimentally obtained value of $E_4$ fairly matches with the value $80$ meV obtained from calculation.

The polarization-resolved photocurrent measurement on the device (geometry: finger width $10\mu$m and gap $10\mu$m; bias voltage=10 V) fabricated on Al$_{0.6}$Ga$_{0.4}$N shows different responsivity spectra $Rc$ and $Rm$ for different in-plane polarization $E\parallel c$ and $E \bot c$ respectively, as shown in Fig.2(b). Inset shows the difference in responsivity ($Rc-Rm$) as a function of wavelength. It shows a peak at $\sim$$265$ nm with peak responsivity of $\sim$$15$\% to the maximum responsivity $Rc$ and FWHM of $\sim$$10$nm. The UV to visible rejection ratio is $10^{2}$. The polarization sensitivity contrast ($Rc/Rm$) is about 1.2. Both the spectra shows cut-off below $280$nm, fulfilling the solar-blind criteria, and making this perhaps the first demonstration polarization sensitive SBUV detectors reported so far.

\begin{figure}[!t]
\centerline{\includegraphics*[width=16cm]{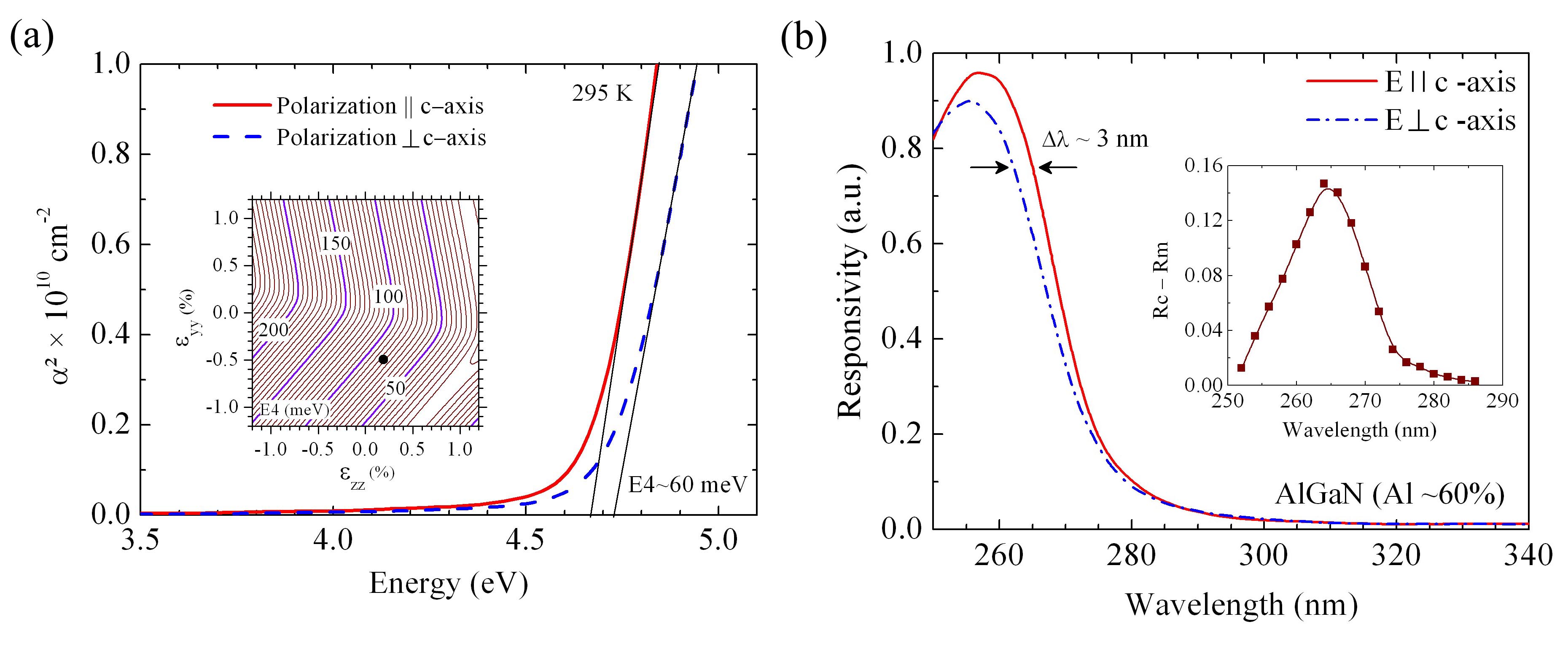}}
\caption{ (a) Optical absorption spectra of $a$-plane Al$_{0.6}$Ga$_{0.4}$N showing difference in bandgap $E_4$$\approx 60$ meV for two different polarizations, Inset: calculated $E_4$ as a function of in-plane strains, black dot represent the strain in our layer for which the calculated value is $\sim 80$ meV (b) Polarization resolved photocurrent measurement for $E\parallel c$ and $E \bot c$ polarization, confirming polarization sensitivity with sharp cut-off below $280$ nm. Inset: different in responsivity as a function of wavelength.}
\end{figure}

In conclusion, we have successfully demonstrated polarization-sensitive SBUV detectors fabricated on non-polar $a$-plane AlGaN. Such devices will be helpful for civil and strategic applications.

\vspace{0.1in}
\noindent\textbf{References:}\\
\noindent\footnotesize{[1] E. Monroy \emph{et al.} Semicond. Sci. Technol. \textbf{18} (2003) R33-R51.}\\
\footnotesize{[2] M.A. Khan \emph{et al.} Jpn. J. Appl. Phys. \textbf{44} (2005) 7191-7206.}\\
\footnotesize{[3] M. Razeghi \emph{et al.} J. Appl. Phys. \textbf{79} (1996) 7433.}\\
\footnotesize{[4] S. Ghosh \emph{et al.} Appl. Phys. Lett. \textbf{90} (2007) 091110.}\\
\footnotesize{[5] A. Navarro \emph{et al.} Appl. Phys. Lett. \textbf{94} (2009) 213512.}\\
\footnotesize{[6] M. R. Laskar, \emph{et al.}  Phys. Stat. Sol. (RRL) \textbf{4}, (2010) 163.}\\
\footnotesize{[7] M. R. Laskar, \emph{et al.}  J. Appl. Phys. \textbf{109}, (2011) 013107.}\\
\footnotesize{[8] J. Bhattacharya \emph{et al.} Phys. Status Solidi B \textbf{246}, 1184 (2009).}\\
\footnotesize{[9] M. R. Laskar, \emph{et al.}  Appl. Phys. Lett. \textbf{98}, (2011) 181108.}\\
\end{document}